\documentstyle[12pt]{article}
\newcommand{\Z}{{\sf Z \!\!\! Z}}

\newcommand{\C}{{\sf C \!\!\! C}}

\makeatletter
\@addtoreset{equation}{section}
\makeatother
 
\title{Quark Confinement in $C$-periodic Cylinders at Temperatures above $T_c$
\footnote{This work is supported in part by funds provided by the U.S.
Department of Energy (D.O.E.) under cooperative research agreement
DE-FC02-94ER40818.}}
\author{K. Holland and U.-J. Wiese \\ \\
Center for Theoretical Physics, \\
Laboratory for Nuclear Science, and Department of Physics \\
Massachusetts Institute of Technology (MIT) \\
Cambridge, Massachusetts 02139, U.S.A. \\ \\
MIT Preprint, CTP 2608 \\ \\}
 
\begin{document}
\maketitle
\begin{abstract} \normalsize
 
Due to the Gauss law, a single quark cannot exist in a periodic volume, while 
it can exist with $C$-periodic boundary conditions. In a $C$-periodic cylinder
of cross section $A = L_x L_y$ and length $L_z \gg L_x, L_y$ containing
deconfined gluons, regions of different high temperature $\Z(3)$ phases are
aligned along the $z$-direction, separated by deconfined-de\-con\-fined 
interfaces. In this geometry, the free energy of a single static quark diverges
in proportion to $L_z$. Hence, paradoxically, the quark is confined, although 
the temperature $T$ is larger than $T_c$. At $T \approx T_c$, the confined phase 
coexists with the three deconfined phases. The deconfined-deconfined interfaces
can be completely or incompletely wet by the confined phase. The free energy of
a quark behaves differently in these two cases. In contrast to claims
in the literature, our results imply that deconfined-deconfined interfaces are 
not Euclidean artifacts, but have observable consequences in a system of hot 
gluons.

\end{abstract}
 
\maketitle
 
\newpage

Let us consider $SU(3)$ gauge fields, $A_\mu(\vec x,t) = i e A^a_\mu(\vec x,t)
\lambda^a$, in four-dimensional Euclidean space-time at finite temperature
$T = 1/\beta$, i.e. with periodic boundary conditions, $A_\mu(\vec x,t + \beta) 
= A_\mu(\vec x,t)$, in the Euclidean time direction. The action,
$S[A_\mu] = \int_0^\beta dt \int d^3x \ (1/2 e^2) \mbox{Tr}
F_{\mu\nu} F_{\mu\nu}$, is defined in terms of the field strength,
$F_{\mu\nu} = \partial_\mu A_\nu - \partial_\nu A_\mu + [A_\mu,A_\nu]$. It is 
invariant under gauge transformations $A_\mu' = g^\dagger (A_\mu + 
\partial_\mu) g$, where the $SU(3)$ matrices, $g(\vec x,t + \beta) = 
g(\vec x,t)$, are also periodic in Euclidean time. The Polyakov loop
\begin{equation}
\Phi(\vec x) = \mbox{Tr}[{\cal P} \exp \int_0^\beta dt \ A_4(\vec x,t)]
\in \C,
\end{equation}
winds around the Euclidean time direction and is constructed from the 
Euclidean time component $A_4$ of the non-Abelian gauge potential. Note that
${\cal P}$ denotes path ordering. The Polyakov loop is a gauge invariant 
complex valued scalar field in three dimensions. Under transformations
\begin{equation}
g(\vec x,t + \beta) = g(\vec x,t) z,
\end{equation}
which are periodic up to a center element $z \in \Z(3) = \{\exp(2 \pi i n/3), \
n = 1,2,3\}$, the Polyakov loop changes into
\begin{equation}
\Phi'(\vec x) = \Phi(\vec x) z,
\end{equation}
while the action remains invariant. The
expectation value $\langle \Phi \rangle = \exp(- \beta F)$ measures 
the free energy, $F$, of a static quark. In the confined phase, $F$ diverges and 
$\langle \Phi \rangle$ vanishes, while in the deconfined phase $F$ is finite
and $\langle \Phi \rangle \neq 0$. Hence, the $\Z(3)$ center symmetry is
spontaneously broken at high temperatures \cite{McL81}. 

In a situation with 
spontaneous symmetry breaking, the physics is sensitive to the spatial boundary 
conditions and to the way in which the infinite volume limit is approached. 
Here, we consider spatial volumes of size $L_x \times L_y \times L_z$ with 
periodic boundary conditions in the $x$- and $y$-directions. If one chooses
periodic boundary conditions also in the $z$-direction, the expectation value
of the Polyakov loop vanishes even in the deconfined phase. This is a
consequence of the Gauss law in a periodic volume \cite{Hil83}. For topological
reasons, a single quark cannot exist in a periodic box, because its center 
electric flux cannot go to infinity. It can only end in an anti-quark such that
the total system is $\Z(3)$ neutral. 

In this letter, we consider $C$-periodic boundary conditions in the 
$z$-direction \cite{Kro91}. When a $C$-periodic field is shifted by $L_z$, it is
replaced by its charge conjugate, i.e. for $C$-periodic gluons
\begin{equation}
A_\mu(\vec x + L_z \vec e_z,t) = A_\mu(\vec x,t)^*.
\end{equation}
Here $*$ denotes complex conjugation. In a $C$-periodic volume, a single quark
can exist, because now its center electric flux can escape to its charge
conjugate partner on the other side of the boundary. Note that the system is
still translationally invariant. The allowed gauge transformations, as well as
the Polyakov loop, also satisfy $C$-periodicity
\begin{equation}
g(\vec x + L_z \vec e_z,t) = g(\vec x,t)^*, \
\Phi(\vec x + L_z \vec e_z) = \Phi(\vec x)^*.
\end{equation}
As a consequence of the boundary conditions, the $\Z(3)$ center symmetry is now
explicitly broken \cite{Wie92}. If one again considers a transformation 
$g(\vec x,t + \beta) = g(\vec x,t) z$, which is periodic, up to a center element
$z \in \Z(3)$, in the Euclidean time direction, one finds
\begin{equation}
g(\vec x,t)^* = g(\vec x,t + \beta)^* z = g(\vec x + L_z \vec e_z,t + \beta) z
= g(\vec x + L_z \vec e_z,t) z^2 = g(\vec x,t)^* z^2.
\end{equation}
Consistency requires $z^2 = 1$ and hence $z = 1$ (because $z \in \Z(3)$). Of
course, in the infinite volume limit, the explicit $\Z(3)$ symmetry breaking
due to the spatial boundary conditions disappears. With $C$-periodic boundary
conditions, $\langle \Phi \rangle$ is always non-zero in a finite volume. In the
confined phase, $\langle \Phi \rangle$ goes to zero in the infinite volume limit,
while it remains finite in the high-temperature deconfined phase. $C$-periodic
boundary conditions are well-suited for studying the free energy of single
quarks, while with periodic boundary conditions a single quark cannot even
exist.

The gluon system is known to have a first order phase transition \cite{Gav89} at a
temperature $T_c$. First, we consider the system in the confined 
phase at temperatures $T < T_c$ in a cylindrical volume with cross section 
$A = L_x L_y$ and length $L_z \gg L_x,L_y$. Diagrammatically, the partition 
function is
\begin{equation}
Z = \begin{picture}(90,15)
\put(0,-4){\line(1,0){90}} \put(0,11){\line(1,0){90}}
\put(0,-4){\line(0,1){15}} \put(90,-4){\line(0,1){15}} \put(43,0){$c$}
\end{picture} \ = \exp(- \beta f_c A L_z),
\end{equation}
where $f_c$ is the temperature-dependent free energy density in the confined
phase. The expectation value of the Polyakov loop (times $Z$), on the other
hand, is given by
\begin{equation}
Z \langle \Phi \rangle = \begin{picture}(90,15)
\put(0,-4){\line(1,0){90}} \put(0,11){\line(1,0){90}}
\put(0,-4){\line(0,1){15}} \put(90,-4){\line(0,1){15}} \put(43,0){$c$}
\put(0,3.5){\line(1,0){38}} \put(52,3.5){\line(1,0){38}}
\end{picture} \ = \exp(- \beta f_c A L_z) \Sigma_0 \exp(- \beta \sigma L_z),
\end{equation}
where $\sigma$ is the string tension, which is again temperature-dependent. The
confining string (denoted by the additional line in the diagram) connects the 
static quark with its anti-quark partner on the other side of the $C$-periodic 
boundary. Hence, the free energy of the quark is given by
\begin{equation}
F = - \frac{1}{\beta} \log \Sigma_0 + \sigma L_z.
\end{equation}
The free energy diverges as $L_z \rightarrow \infty$, indicating that the quark
is confined.

Now let us consider the deconfined phase at temperatures $T > T_c$, where three
distinct deconfined phases coexist. They are distinguished by different values 
for the Polyakov loop and are related to each other by $\Z(3)$ transformations.
The expectation values of the Polyakov loop in the three phases are
\begin{equation}
\Phi^{(1)} = (\Phi_0,0), \ 
\Phi^{(2)} = (- \frac{1}{2} \Phi_0,\frac{\sqrt{3}}{2} \Phi_0), \
\Phi^{(3)} = (- \frac{1}{2} \Phi_0,- \frac{\sqrt{3}}{2} \Phi_0).
\end{equation}
In a cylindrical volume, a typical configuration consists of several bulk 
phases, aligned along the $z$-direction, separated by deconfined-deconfined 
interfaces. These interfaces cost free energy $F$ proportional to their area 
$A$, such that their interface tension is given by $\alpha_{dd} = F/A$.
What matters in the following is the cylindrical shape, not the magnitude of
the volume. In fact, our cylinders can be of macroscopic size. The expectation 
value of the Polyakov loop in a cylindrical volume can be calculated from a 
dilute gas of interfaces \cite{Gro93}. The interface expansion of the partition 
function can be viewed as
\begin{equation}
Z = \begin{picture}(90,15)
\put(0,-4){\line(1,0){90}} \put(0,11){\line(1,0){90}}
\put(0,-4){\line(0,1){15}} \put(90,-4){\line(0,1){15}} \put(40,0){$d_1$}
\end{picture} \ + \
\begin{picture}(90,15)
\put(0,-4){\line(1,0){90}} \put(0,11){\line(1,0){90}}
\put(0,-4){\line(0,1){15}} \put(90,-4){\line(0,1){15}} 
\put(45,-4){\line(0,1){15}} \put(18,0){$d_2$} \put(63,0){$d_3$} 
\end{picture} \ + \
\begin{picture}(90,15)
\put(0,-4){\line(1,0){90}} \put(0,11){\line(1,0){90}}
\put(0,-4){\line(0,1){15}} \put(90,-4){\line(0,1){15}}
\put(45,-4){\line(0,1){15}} \put(18,0){$d_3$} \put(63,0){$d_2$}
\end{picture} \ + ...
\end{equation}
The first term has no interfaces and thus the whole cylinder is filled with
deconfined phase $d_1$ only. An entire volume filled with either phase $d_2$ or
$d_3$ would not satisfy the boundary conditions. The second and third terms 
have one interface separating phases $d_2$ and $d_3$. Here, $C$-periodic 
boundary conditions exclude phase $d_1$. The sum of the diagrammatic terms is 
given by
\begin{eqnarray}
Z&=&\exp(- \beta f_d A L_z) \nonumber \\
&+& 2 \int_0^{L_z} dz \ \exp(- \beta f_d A z)
\gamma \exp(- \beta \alpha_{dd} A) \exp(- \beta f_d A (L_z - z)) + ... 
\nonumber \\
&=&\exp(- \beta f_d A L_z)
[1 + 2 \gamma \exp(- \beta \alpha_{dd} A) L_z + ...].
\end{eqnarray}
The first term is the Boltzmann weight of deconfined phase of volume $A L_z$
with a free energy density $f_d$. The second term contains two of these bulk
Boltzmann factors separated by an interface contribution 
$\gamma \exp(- \beta \alpha_{dd} A)$, where $\gamma$ is a factor resulting from
capillary wave fluctuations of the interface. Note that in three dimensions,
$\gamma$ is to leading order independent of the area $A$ \cite{Bre85}. In the 
above expression, we have integrated over all possible locations $z$ of the 
interface. It is straightforward to sum the interface expansion to all orders, 
giving
\begin{equation}
Z = \exp(- \beta f_d A L_z + 2 \gamma \exp(- \beta \alpha_{dd} A) L_z).
\end{equation}
In exactly the same way, the expectation value of the Polyakov loop is given by
\begin{eqnarray}
Z \langle \Phi \rangle&=&\exp(- \beta f_d A L_z)
\{\Phi^{(1)} + \int_0^{L_z} dz \ \gamma \exp(- \beta \alpha_{dd} A) \nonumber \\
&\times&\frac{1}{L_z}
[\Phi^{(2)} z + \Phi^{(3)}(L_z - z) + \Phi^{(3)} z + \Phi^{(2)}(L_z - z)]
+ ...\} \nonumber \\
&=&\Phi_0 \exp(- \beta f_d A L_z)
\{1 - \gamma \exp(- \beta \alpha_{dd} A) L_z + ...\}.
\end{eqnarray}
Summing the whole series and dividing by $Z$, one obtains
\begin{equation}
\langle \Phi \rangle = \Phi_0 \exp(- 3 \gamma \exp(- \beta \alpha_{dd} A) L_z).
\end{equation}
Note that $\langle \Phi \rangle$ is real, as it should be. The free energy of a 
static quark in a $C$-periodic cylinder is therefore given by
\begin{equation}
F = - \frac{1}{\beta} \log \Phi_0 + 
\frac{3 \gamma}{\beta} \exp(- \beta \alpha_{dd} A) L_z.
\end{equation}
This result is counter intuitive. Although we are in the deconfined phase, the
quark's free energy diverges in the limit $L_z \rightarrow \infty$, as long as
the cross section $A$ of the cylinder remains fixed. This is the behavior one
typically associates with confinement. In fact, 
\begin{equation}
\sigma' = \frac{3 \gamma}{\beta} \exp(- \beta \alpha_{dd} A)
\end{equation}
plays the role of the ``string tension'', even though there is no physical 
string that connects the quark with its anti-quark partner on the other side of
the $C$-periodic boundary. Confinement in $C$-periodic cylinders arises because
disorder due to many differently oriented deconfined phases destroys the 
correlations of center electric flux between quark and anti-quark.

Of course, this paradoxical confinement mechanism is due to the cylindrical
geometry and the specific boundary conditions. Had we chosen $C$-periodic 
boundary conditions in all directions, the deconfined phases $d_2$ and $d_3$
would be exponentially suppressed, so that the entire volume would be filled 
with phase $d_1$ only. In that case, the free energy of a static quark is 
$F = - (1/\beta) \log \Phi_0$, which does not diverge in the infinite volume 
limit, as expected. Had we worked in a cubic volume, a typical configuration 
would have no interfaces, the whole volume would be filled with deconfined 
phase $d_1$ and again the free energy of a static quark would be 
$F = - (1/\beta) \log \Phi_0$. Finally, note that, in a cylindrical volume, even
the static Coulomb potential is linearly rising with a ``string tension''
$e^2/A$. Due to Debye screening, this trivial confinement effect is absent
in the gluon plasma.

In contrast to claims in the literature \cite{Smi94}, our result 
implies that de\-con\-fined-deconfined interfaces are more than just Euclidean 
field configurations. In fact, they can lead to a divergence of the free energy
of a static quark and thus they have physically observable consequences. Of 
course, the issue is somewhat academic. First of all, the existence of three 
distinct deconfined phases relies on the $\Z(3)$ symmetry, which only exists 
in a pure gluon system --- not in the real world with light dynamical quarks. 
Secondly, the effect is due to the cylindrical geometry and our choice of 
boundary conditions. Even though it is not very realistic, this set-up 
describes a perfectly well-defined Gedanken-experiment, demonstrating the 
physical reality of deconfined-deconfined interfaces.

Let us now investigate how the paradoxical confinement mechanism turns into the
ordinary one as we lower the temperature to $T \approx T_c$. Since the phase
transition is of first order, around $T_c$, the three deconfined phases  
coexist with the confined phase. Then, there are also confined-deconfined 
interfaces with an interface tension $\alpha_{cd}$. There are two possible scenarios
--- complete or incomplete wetting \cite{Fre89,Tra92}. If $\alpha_{dd} = 
2 \alpha_{cd}$ at $T = T_c$, a deconfined-deconfined interface is unstable and 
splits into a pair of confined-deconfined interfaces. A film of confined phase 
grows to macroscopic size and completely wets the deconfined-deconfined interface. 
Note that $\alpha_{dd} > 2 \alpha_{cd}$ is not possible in thermal equilibrium. 
Thus, the condition for complete wetting does not require fine-tuning. The 
alternative is incomplete wetting, which corresponds to $\alpha_{dd} 
< 2 \alpha_{cd}$. In that case, both confined-deconfined and 
deconfined-deconfined interfaces are stable. Numerical simulations indicate 
that complete wetting is realized in the gluon system \cite{Kaj90,Gro93}.

Assuming complete wetting, the partition function takes the form
\begin{eqnarray}
Z&=&\begin{picture}(90,15)
\put(0,-4){\line(1,0){90}} \put(0,11){\line(1,0){90}}
\put(0,-4){\line(0,1){15}} \put(90,-4){\line(0,1){15}} \put(43,0){$c$}
\end{picture} \ + \
\begin{picture}(90,15)
\put(0,-4){\line(1,0){90}} \put(0,11){\line(1,0){90}}
\put(0,-4){\line(0,1){15}} \put(90,-4){\line(0,1){15}} \put(40,0){$d_1$}
\end{picture} \nonumber \\ \ \nonumber \\
&+&\sum_i \big\{ \ \begin{picture}(90,15)
\put(0,-4){\line(1,0){90}} \put(0,11){\line(1,0){90}}
\put(0,-4){\line(0,1){15}} \put(90,-4){\line(0,1){15}}
\put(30,-4){\line(0,1){15}} \put(60,-4){\line(0,1){15}}
\put(13,0){$c$} \put(40,0){$d_i$} \put(73,0){$c$} 
\end{picture} \ + \
\begin{picture}(90,15)
\put(0,-4){\line(1,0){90}} \put(0,11){\line(1,0){90}}
\put(0,-4){\line(0,1){15}} \put(90,-4){\line(0,1){15}}
\put(30,-4){\line(0,1){15}} \put(60,-4){\line(0,1){15}}
\put(10,0){$d_i$} \put(43,0){$c$} \put(70,0){$d^*_i$}
\end{picture} \ \big\} \ + ...
\end{eqnarray}
The sum over $i$ extends over the three deconfined phases and $d^*_i$ denotes
the charge-conjugate of $d_i$. For example, $d^*_1 = d_1$ and $d^*_2 = d_3$.
Note that due to complete wetting, one always has an even number of interfaces.
The diagrammatic terms from above give
\begin{eqnarray}
Z&=&\exp(- \beta f_c A L_z) + \exp(- \beta f_d A L_z) \nonumber \\
&+&3 \int_0^{L_z} dz_1 \int_{z_1}^{L_z} dz_2 \ \big\{\exp(- \beta f_c A z_1)
\delta \exp(- \beta \alpha_{cd} A) \exp(- \beta f_d A (z_2 - z_1)) \nonumber \\
&\times&\delta \exp(- \beta \alpha_{cd} A) \exp(- \beta f_c A (L_z - z_2)) \
+ \ \exp(- \beta f_d A z_1) \delta \exp(- \beta \alpha_{cd} A) \nonumber \\
&\times&\exp(- \beta f_c A (z_2 - z_1)) \delta \exp(- \beta \alpha_{cd} A) 
\exp(- \beta f_d A (L_z - z_2))\big\} + ...
\end{eqnarray}
Again, there is a factor for every bulk phase and for every interface. The
capillary wave fluctuations of a confined-deconfined interface are characterized
by the constant $\delta$. Note that now we integrate over the positions $z_1$ 
and $z_2$ of the two interfaces. Similarly, the diagrammatic terms for the
expectation value of the Polyakov loop are
\begin{eqnarray}
Z \langle \Phi \rangle&=&\begin{picture}(90,15)
\put(0,-4){\line(1,0){90}} \put(0,11){\line(1,0){90}}
\put(0,-4){\line(0,1){15}} \put(90,-4){\line(0,1){15}} \put(43,0){$c$}
\put(0,3.5){\line(1,0){38}} \put(52,3.5){\line(1,0){38}} \end{picture} \ + \
\begin{picture}(90,15)
\put(0,-4){\line(1,0){90}} \put(0,11){\line(1,0){90}}
\put(0,-4){\line(0,1){15}} \put(90,-4){\line(0,1){15}} \put(40,0){$d_1$}
\end{picture} \nonumber \\ \ \nonumber \\
&+&\sum_i \big\{ \ \begin{picture}(90,15)
\put(0,-4){\line(1,0){90}} \put(0,11){\line(1,0){90}}
\put(0,-4){\line(0,1){15}} \put(90,-4){\line(0,1){15}}
\put(30,-4){\line(0,1){15}} \put(60,-4){\line(0,1){15}}
\put(13,0){$c$} \put(40,0){$d_i$} \put(73,0){$c$} 
\put(0,3.5){\line(1,0){8}} \put(22,3.5){\line(1,0){8}}
\put(60,3.5){\line(1,0){8}} \put(82,3.5){\line(1,0){8}} 
\end{picture} \ + \
\begin{picture}(90,15)
\put(0,-4){\line(1,0){90}} \put(0,11){\line(1,0){90}}
\put(0,-4){\line(0,1){15}} \put(90,-4){\line(0,1){15}}
\put(30,-4){\line(0,1){15}} \put(60,-4){\line(0,1){15}}
\put(10,0){$d_i$} \put(43,0){$c$} \put(70,0){$d^*_i$}
\put(30,3.5){\line(1,0){8}} \put(52,3.5){\line(1,0){8}}
\end{picture} \ \big\} \ + ...
\end{eqnarray}
The center electric flux connecting the quark with its $C$-periodic partner is 
constricted into a tube when it passes through a region of confined phase.
Summing the series for $Z$ and $Z \langle \Phi \rangle$ as we did before yields
\begin{equation}
\langle \Phi \rangle = \frac{\Phi_0 \exp(\beta x L_z) + 
\Sigma_0 \exp(- \beta (x + \sigma) L_z)}
{2 \cosh[\beta L_z \sqrt{x^2 + (3 \delta^2/\beta^2) 
\exp(- 2 \beta \alpha_{cd} A)}]}.
\end{equation}
Here
\begin{equation}
x = \frac{1}{2} (f_c - f_d) A
\end{equation}
measures the free energy difference between confined and deconfined phases.
If $f_d < f_c + \sigma/A$, the resulting ``string tension'' is given by
\begin{equation}
\sigma' = \sqrt{x^2 + (3 \delta^2/\beta^2) \exp(- 2 \beta \alpha_{cd} A)} - x,
\end{equation}
which is independent of the string tension $\sigma$ in the confined phase.
In particular, at the finite volume critical point, i.e. at $f_d = f_c$, we have
\begin{equation}
\sigma' = \frac{\sqrt{3} \delta}{\beta} \exp(- \beta \alpha_{cd} A),
\end{equation}
For $f_d > f_c + \sigma/A$, on the other hand,
\begin{equation}
\sigma' = \sigma +
\sqrt{x^2 + (3 \delta^2/\beta^2) \exp(- 2 \beta \alpha_{cd} A)} + x.
\end{equation}
Note that now $x<0$ and hence, as expected, $\sigma'$ reduces to $\sigma$ in 
the large $A$ limit.

For completeness, let us also discuss the incomplete wetting case, even though
it seems not to be realized in the gluon system. Diagrammatically, the
partition function then takes the form
\begin{eqnarray}
Z&=&\begin{picture}(90,15)
\put(0,-4){\line(1,0){90}} \put(0,11){\line(1,0){90}}
\put(0,-4){\line(0,1){15}} \put(90,-4){\line(0,1){15}} \put(43,0){$c$}
\end{picture} \ + \
\begin{picture}(90,15)
\put(0,-4){\line(1,0){90}} \put(0,11){\line(1,0){90}}
\put(0,-4){\line(0,1){15}} \put(90,-4){\line(0,1){15}} \put(40,0){$d_1$}
\end{picture} \ + \
\begin{picture}(90,15)
\put(0,-4){\line(1,0){90}} \put(0,11){\line(1,0){90}}
\put(0,-4){\line(0,1){15}} \put(90,-4){\line(0,1){15}} 
\put(45,-4){\line(0,1){15}} \put(18,0){$d_2$} \put(63,0){$d_3$}
\end{picture} \nonumber \\ \ \nonumber \\
&+&\begin{picture}(90,15)
\put(0,-4){\line(1,0){90}} \put(0,11){\line(1,0){90}}
\put(0,-4){\line(0,1){15}} \put(90,-4){\line(0,1){15}}
\put(45,-4){\line(0,1){15}} \put(18,0){$d_3$} \put(63,0){$d_2$} 
\end{picture} \ + \ 
\sum_i \big\{ \ \begin{picture}(90,15)
\put(0,-4){\line(1,0){90}} \put(0,11){\line(1,0){90}}
\put(0,-4){\line(0,1){15}} \put(90,-4){\line(0,1){15}}
\put(30,-4){\line(0,1){15}} \put(60,-4){\line(0,1){15}}
\put(13,0){$c$} \put(40,0){$d_i$} \put(73,0){$c$} 
\end{picture} \ + \
\begin{picture}(90,15)
\put(0,-4){\line(1,0){90}} \put(0,11){\line(1,0){90}}
\put(0,-4){\line(0,1){15}} \put(90,-4){\line(0,1){15}}
\put(30,-4){\line(0,1){15}} \put(60,-4){\line(0,1){15}}
\put(10,0){$d_i$} \put(43,0){$c$} \put(70,0){$d^*_i$}
\end{picture} \ \big\} \nonumber \\ \ \nonumber \\
&+&\sum_{i,j} \ \begin{picture}(90,15)
\put(0,-4){\line(1,0){90}} \put(0,11){\line(1,0){90}}
\put(0,-4){\line(0,1){15}} \put(90,-4){\line(0,1){15}}
\put(30,-4){\line(0,1){15}} \put(60,-4){\line(0,1){15}}
\put(10,0){$d_i$} \put(40,0){$d_j$} \put(70,0){$d^*_i$}
\end{picture} \ + ...
\end{eqnarray}
Now we also have deconfined-deconfined interfaces. The sum in the last term is 
restricted to $d_j \neq d_i, d^*_i$. In complete analogy to the previous 
calculations, one obtains
\begin{eqnarray}
\langle \Phi \rangle&=&\frac{\Phi_0 
\exp(\beta (x - (\gamma/\beta) \exp(- \beta \alpha_{dd} A))  L_z) + 
\Sigma_0 \exp(- \beta (x + \sigma) L_z)}
{2 \cosh[\beta L_z \sqrt{(x + (\gamma/\beta) \exp(- \beta \alpha_{dd} A))^2 + 
(3 \delta^2/\beta^2) \exp(- 2 \beta \alpha_{cd} A)}]} \nonumber \\ 
&\times&\exp(- \gamma \exp(- \beta \alpha_{dd} A) L_z).
\end{eqnarray}
It is straightforward to extract the ``string tension'' $\sigma'$ and examine 
the limiting cases as we did for complete wetting.

In conclusion, we have found a paradoxical mechanism for confinement that works
even deep in the deconfined phase, provided the Universe is a cylinder and has
specific boundary conditions. Ignoring subtle infra-red effects related to the
boundary conditions in a situation with spontaneously broken symmetry has led
other authors to conclude that deconfined-deconfined interfaces are just
artifacts of the Euclidean description. Our calculation relates the presence of
interfaces to the free energy of a static quark and thus shows that
deconfined-deconfined interfaces have physically observable consequences. Even
though an experimentalist cannot choose the boundary conditions of the 
Universe, a lattice field theorist, performing a numerical simulation of the 
gluon system, can. In fact, one can use our results to extract the string 
tension $\sigma$, as well as the interface tensions $\alpha_{cd}$ and $\alpha_{dd}$,
from simple lattice measurements of a single Polyakov loop. This may turn out 
to be technically easier than applying standard methods.

\section*{Acknowledgements}

We are indebted to R. Brower, S. Chandrasekharan and S. Bashinsky for very 
interesting discussions. One of the authors (U.-J. W.) wishes to thank the 
theory group of the Weizmann Institute in Rehovot, where part of this work was 
done, for its hospitality, and the A. P. Sloan foundation for its support.


\begin{thebibliography}{10}

\bibitem{McL81}
L. D. McLerran and B. Svetitsky, Phys. Rev. D24 (1981) 450.

\bibitem{Hil83}
E. Hilf and L. Polley, Phys. Lett. B131 (1983) 412.

\bibitem{Kro91}
A. S. Kronfeld and U.-J. Wiese, Nucl. Phys. B357 (1991) 521.

\bibitem{Wie92}
U.-J. Wiese, Nucl. Phys. B375 (1992) 45.

\bibitem{Gav89}
R. V. Gavai, F. Karsch and B. Petersson, Nucl. Phys. B322 (1989) 738; \\
M. Fukugita, M. Okawa and U. Ukawa, Phys. Rev. Lett. 63 (1989) 1768; \\
N. A. Alves, B. A. Berg and S. Sanielevici, Phys. Rev. Lett. 64 (1990) 3107.

\bibitem{Gro93}
B. Grossmann, M. L. Laursen, T. Trappenberg and U.-J. Wiese, Nucl. Phys. B396
(1993) 584.

\bibitem{Bre85}
V. Privman and M. E. Fisher, J. Stat. Phys. 33 (1983) 385; \\
E. Brezin and J. Zinn-Justin, Nucl. Phys. B257 [FS14] (1985) 867.

\bibitem{Smi94}
A. V. Smilga, Ann. Phys. 234 (1994) 1; \\
J. Kiskis, hep-lat/9510029; \\
T. H. Hansson, H. B. Nielsen and I. Zahed, Nucl. Phys. B451 (1995) 162.

\bibitem{Fre89}
Z. Frei and A. Patk\'{o}s, Phys. Lett. B229 (1989) 102.

\bibitem{Tra92}
T. Trappenberg and U.-J. Wiese, Nucl. Phys. B372 (1992) 703.

\bibitem{Kaj90}
K. Kajantie, L. K\"arkk\"ainen and K. Rummukainen, Nucl. Phys. B333 (1990) 100;
Nucl. Phys. B357 (1991) 693.


\end{thebibliography}
\end{document}